\documentclass[runningheads]{svmult}

\usepackage{makeidx}   
\usepackage{graphicx}  
\usepackage{subeqnar}  
\usepackage{multicol}  
\usepackage{physprbb}  
\makeindex             

\newcommand{\LCDM}{\mbox{${\rm \Lambda}$CDM}}
\newcommand{\msun}{\mbox{${\rm M}_{\odot}$}}
\def\spose#1{\hbox to 0pt{#1\hss}}
\def\la{\mathrel{\spose{\lower 3pt\hbox{$\mathchar"218$}}
     \raise 2.0pt\hbox{$\mathchar"13C$}}}
\def\ga{\mathrel{\spose{\lower 3pt\hbox{$\mathchar"218$}}
     \raise 2.0pt\hbox{$\mathchar"13E$}}}

\begin{document}
\title*{Modeling the Multi-wavelength Universe: the assembly of massive galaxies}
\toctitle{Modeling the Multi-wavelength Universe: the assembly of massive galaxies}
\titlerunning{Modeling the Multi-wavelength Universe}
\author{Rachel S. Somerville}
\authorrunning{R.S. Somerville}
\institute{Space Telescope Science Institute}

\maketitle              

\begin{abstract}
Pan-chromatic galaxy surveys are providing tightening constraints on
the global mass assembly history, and high-resolution imaging of large
fields is telling us when and where stars were formed. How well are
state-of-the-art hierarchical galaxy formation models currently doing
at reproducing these observations? I present results here that suggest
that hierarchical models are doing quite well at reproducing the
\emph{global} star formation and stellar mass assembly history
obtained from galaxies selected in the optical and Near IR. However,
the same models fail to reproduce two very important populations at
high redshift: quiescent red spheroids and vigorously star-forming,
dust-enshrouded starbursts. This mismatch carries important lessons
about how star formation is triggered and regulated in early galaxies,
and may force us to consider new ideas about the formation of massive
spheroids.
\end{abstract}

\section{Tracing Galaxy Assembly}

Probing the high-mass end of the galaxy mass function at any epoch
provides a particularly strong test of theories of galaxy formation. A
zoo of apparently massive objects at high redshift has been
discovered in recent years, in a broad range of wavelengths and
utilizing a variety of selection techniques. For example, at this
meeting, we heard about massive galaxies at $z\sim1$ in the
COMBO-17/GEMS survey (Bell, Rix), VIMOS (Le Fevre), and DEEP (Koo,
Newman), near IR selected galaxies and Extremely Red Objects (EROs) at
$z\sim1$--2 (Fontana, Daddi, Drory, Cimatti), color-selected galaxies
in the spectroscopic redshift `desert' at $z\sim 1.5$-2.5 (Steidel,
Chen), and sub-mm and mm selected galaxies at $z\sim 2.5$ (Chapman,
Bertoldi, Genzel). The challenge posed to us by the multi-wavelength
universe is to unify this zoo of objects into a coherent picture of
galaxy evolution.

Another exciting observational development is the imaging of
relatively large fields with the Advanced Camera for Surveys (ACS) on
the Hubble Space Telescope (HST), combined with multi-wavelength
ground-based observations. These new surveys, such as GOODS (Great
Observatories Origins Deep Survey), GEMS (Galaxy Evolution from
Morphology and SEDs), and COSMOS, will for the first time allow us to
study the connection between galaxy morphology (through structural
parameters such as bulge-to-disk ratio) and stellar populations
(through broad band colors and spectroscopy) for a statistically
meaningful sample, over a large range of cosmic time. We can then go
beyond a global census of star formation or the build-up of stellar
mass, to understand \emph{where and how} stars were formed, and
perhaps which processes stimulated or regulated star formation. In
this paper, I shall attempt to briefly summarize the status of our
theoretical understanding of some of these observational results and
what we may learn about galaxy formation from them.

\section{Do Massive Galaxies at High Redshift Pose a `Crisis' for CDM?}
\begin{figure}[t]
\begin{center}
\includegraphics[width=0.9\textwidth]{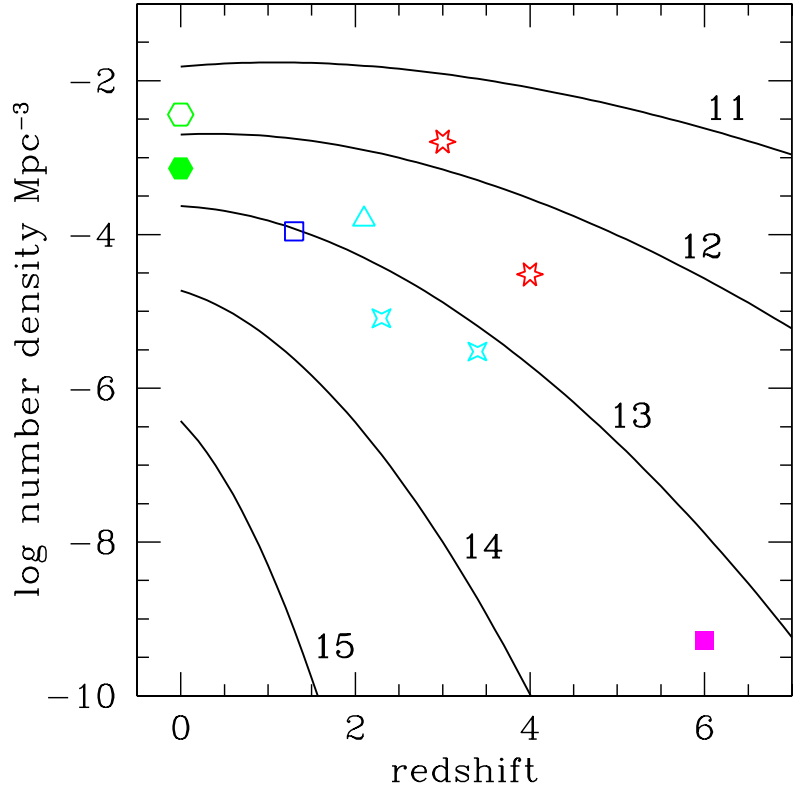}
\end{center}
\caption[]{Curves show the cumulative number density of dark matter
halos more massive than $10^{11}$, $10^{12}$, $10^{13}$, $10^{14}$,
and $10^{15}$ \msun (from top to bottom, as labeled), as a function
of redshift. Points show the estimated comoving number densities of
several observed populations, as follows. Hexagons: galaxies with
stellar masses greater than $10^{11}\msun$ (lower point) and
$2.5\times10^{10}\msun$ (higher point), obtained by integrating the
stellar mass function of \protect\cite{bell:03} from SDSS+2MASS; open
square: Extremely Red Objects \protect\cite{moustakas:04}; ppen
triangle: K20 galaxies \protect\cite{daddi:03}; crosses: sub-mm
galaxies \protect\cite{chapman:03}; six pointed stars: Lyman break
galaxies \protect\cite{steidel:99}, filled square: quasars
\protect\cite{fan:01}. For all populations, \LCDM\ predicts that there
are enough dark matter halos massive enough to plausibly host the
observed objects (see text).}
\label{fig:numz}
\end{figure}

It has become a familiar story: in the Cold Dark Matter (CDM) model of
structure formation, small mass objects form first, and larger mass
objects form hierarchically through mergers and accretion. Therefore,
measurements of the number density of massive structures at high
redshift pose \emph{potentially} strong constraints on this class of
models. The extent to which this potential is realized, however,
depends on how massive and how common are the objects at any given
epoch. Moreover, such arguments also require making a connection
between an observed population and the dark matter halos in which they
are expected to reside. For most, if not all, of the populations
mentioned above, this is far from straightforward. It is a common
perception that the detection of massive galaxies at high redshift is
a serious problem for CDM. It is interesting to take a step back for a
moment from the complexities of modeling gas physics, star formation,
feedback, radiative transfer, stellar populations, etc., and to
simply ask whether the `concordance' \LCDM\ model produces enough
massive \emph{dark matter halos} to plausibly host the objects that
have been detected.

This simple exercise is carried out in Fig.~\ref{fig:numz}, which
shows the predicted cumulative comoving number density of dark matter
halos above a given mass, from $10^{11}$--$10^{15}\, \msun$ as labeled,
as a function of redshift\footnote{All results presented in this paper
assume the `concordance' \LCDM\ cosmology: $\Omega_m=0.3$,
$\Omega_{\Lambda}=0.7$, $H_0=70$ km/s/Mpc, $\sigma_8$=0.9.}.  These
were obtained from the Sheth-Tormen modified Press-Schechter model
\cite{sheth-tormen}. Also shown are the estimated comoving number
densities of various observed populations, as described in the figure
caption.  It has been claimed that the sub-mm galaxies, EROs and K20
$z\sim 2$ objects have stellar masses of a few times $10^{11} \msun$
\cite{chapman:03,genzel:venice,moustakas:pc,daddi:03}. Making the
plausible assumptions that about fifteen percent of the total mass is
in the form of baryons, and a few tenths of the baryons are in stars,
one can easily accommodate the observed numbers of objects within
suitably massive dark matter halos of $\sim 10^{13} \msun$. The Lyman
break galaxies, which are more numerous but probably have stellar
masses about an order of magnitude smaller (a few times $10^{10}
\msun$; \cite{shapley,papovich}) are also easily accommodated within
halos of a few times $10^{11} \msun$. Even the $z\sim 6$ quasars
detected by SDSS \cite{fan:01}, which must harbor black holes of $\sim
10^9 \msun$ if they are Eddington limited, can be easily accommodated
within $10^{13} \msun$ halos (see \cite{bsf04} for more detailed
modeling of these objects). So far, none of the observed populations
poses a \emph{fundamental} problem for concordance $\LCDM$. Had the
observed number density of $10^{11} \msun$ galaxies at $z\sim2$ been
an order of magnitude higher than the current estimates, theorists
would have good cause to squirm. As it is, these results suggest that
there were fewer massive galaxies in the past, in qualitative
agreement with hierarchical structure formation. The challenge now
lies in understanding how, where and why the stars and dust that make
these objects visible to us are formed. This requires getting our
hands a bit more dirty.

\section{The Global Mass Assembly History}
\begin{figure}[t]
\begin{center}
\includegraphics[width=0.9\textwidth]{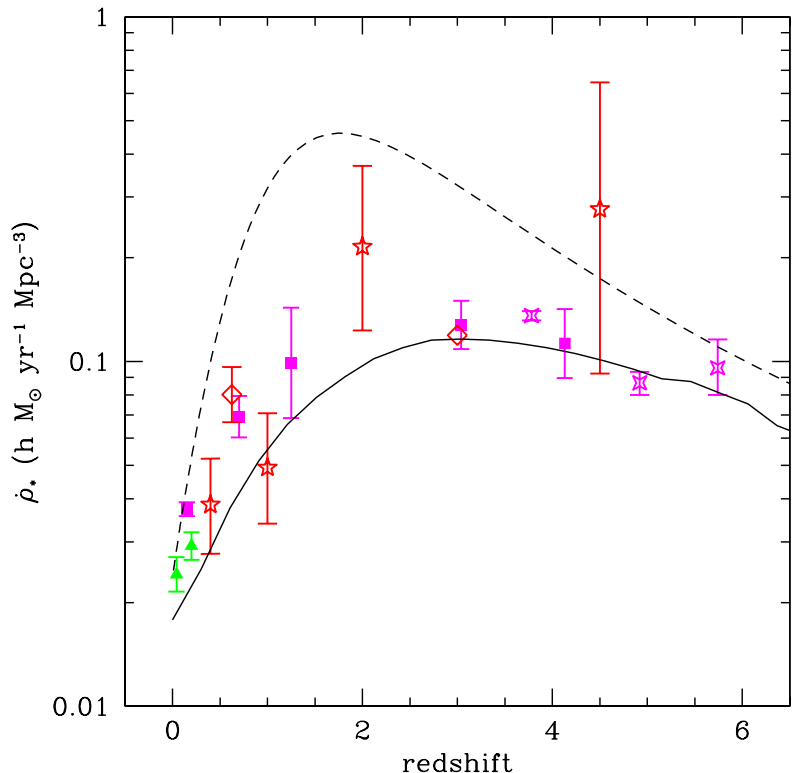}
\end{center}
\caption[]{The global star formation history. The solid curve shows
the prediction from a semi-analytic model (similar to the
`accelerated quiescent' model of \protect\cite{spf}). The dashed line
shows the model from \protect\cite{blain:99}, which was designed to
reproduce the counts of bright sub-mm galaxies and the Far-IR
extragalactic background. Triangles are observational estimates based
on H$\alpha$, squares are based on rest UV, and open diamonds are
based on mid-IR or sub-mm observations (see \protect\cite{spf} for
references). The crosses are new measurements, based on rest UV, from
\protect\cite{giavalisco:04}. Star symbols show measurements based on
sub-mm observations from \protect\cite{barger}. All optical/UV
measurements are corrected for dust extinction and converted to a
$\LCDM$ cosmology where necessary, as described in \protect\cite{spf}.
}
\label{fig:sfh}
\end{figure}

\begin{figure}[t]
\begin{center}
\includegraphics[width=0.9\textwidth]{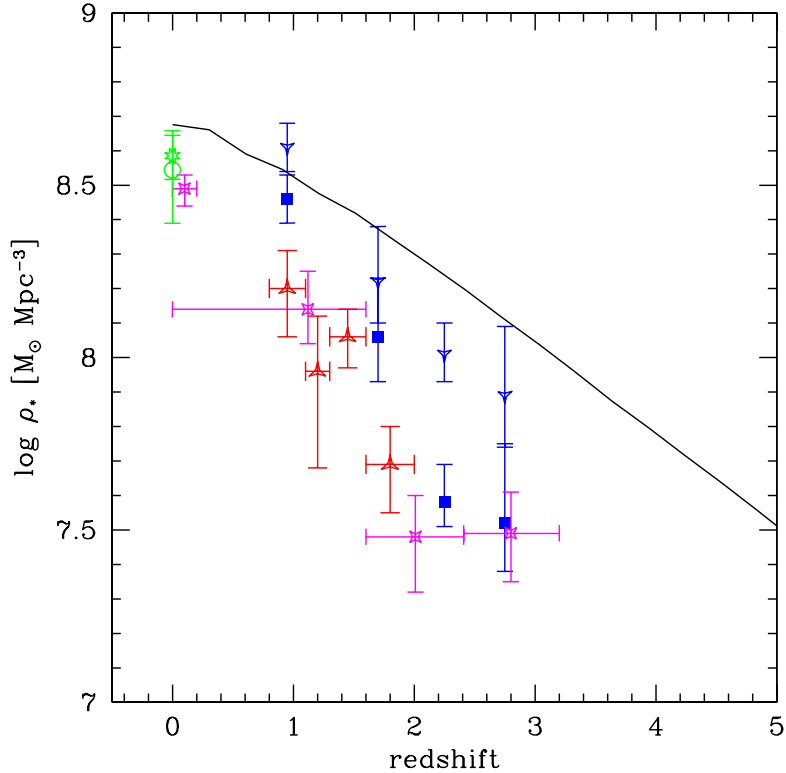}
\end{center}
\caption[]{Global stellar mass assembly history.  The solid line shows
the total comoving density of stellar mass as a function of redshift,
from the same semi-analytic model shown in Fig.~\ref{fig:sfh}. Symbols
show observational estimates of the stellar mass density at various
redshifts from summing the mass in established galaxies. Open circle
and many-pointed star: local estimates from \protect\cite{cole:01} and
\protect\cite{bell:03}. Filled squares: \protect\cite{dpfb};
downward-pointing triangles: `maximum mass' estimates from
\protect\cite{dpfb}; crosses: \protect\cite{rudnick}; upward-pointing
arrows: \protect\cite{gdds}. The Dickinson et al. and $z=0$ points are
meant to represent \emph{total} masses, while the Rudnick et al. and
Glazebrook et al. points are integrated above a fixed rest-frame
luminosity or stellar mass, respectively. }
\label{fig:mah}
\end{figure}

There are several distinct methods of tracing the global history of
star formation and the build-up of stellar mass in the Universe
observationally. One is to measure star formation rates by directly
counting actively star forming galaxies via various tracers (e.g.,
\cite{lilly,madau}). Another is to detect established galaxies at
different redshifts, estimate their stellar masses from their colors
or spectra, and sum the masses up to obtain a global average
\cite{dpfb,rudnick,gdds}. Yet a third approach is to measure the ages
of stellar populations and use galactic `archeology' to deduce
something about star formation at earlier times
(e.g. \cite{renzini}). It is useful to compare the results obtained
from all three methods, as the attendant errors and selection effects
are different.

Fig.~\ref{fig:sfh} shows a compilation of measurements of the star
formation history of the Universe from $z\sim6$ to the present,
including the recent results from GOODS \cite{giavalisco:04}. The
solid curve shows the integrated star formation rate density predicted
by a fairly standard semi-analytic model (similar to the `accelerated
quiescent' model of \cite{spf}). It was shown in \cite{spf} that this
model produced good agreement with existing (dust-corrected)
observational estimates of the star formation history to $z\sim4$; it
is encouraging that the same model agrees almost uncannily well with
the new measurements out to $z\sim6$ from \cite{giavalisco:04}. The
dashed curve is the updated model of \cite{blain:99}, which I shall
return to later. Let us now turn to Fig.~\ref{fig:mah}, which shows
the total \emph{stellar mass density} as a function of redshift. The
solid line is the mass density of all stars from the same
semi-analytic model shown in Fig.~\ref{fig:sfh}. Symbols show
observational estimates, as described in the figure caption.

Obviously, in an ideal world, and neglecting recycling from stellar
mass loss, this curve should be the integral of the curve shown in
Fig.~\ref{fig:sfh}. For the semi-analytic model, this is indeed the
case. Given that the semi-analytic model is quite a good fit to the
observational results shown in Fig.~\ref{fig:sfh}, we would obtain
something quite similar if we simply integrated a curve fit to the
observed star formation history. However, although if anything the SFR
produced by the semi-analytic model is a bit on the low side at low
redshift (which dominates the integral), the total mass produced by
the present day is on the high side. At $z\ga1$, the observational
estimates of the total mass in established systems are more than 50
percent lower than the model, or than the integrated star formation
curve.  This could be due to several factors. One possibility is that
the IMF changes with time (for example, is more top-heavy at early
times). Another is that the total star formation rate has been
overestimated, perhaps because of dust corrections, or completeness
corrections. A third is that galaxies are missing from the stellar
mass census (\cite{idzi} show that this is likely to be the case at
high redshift) or that the stellar masses are systematically
underestimated.

This shows that a \LCDM-based galaxy formation model has no difficulty
producing enough early stars or star formation
\emph{globally}. Similar results have been found based on hydrodynamic
N-body simulations by e.g. \cite{springel}. Similarly, the model
predicts that roughly half of the present day stellar mass was in
place by $z\sim2$, and seventy percent by $z\sim1$, implying that
there should be enough old stars ($\ga 8-10$ Gyr) at $z=0$ to satisfy
the `fossil evidence' in present day galaxies.

\section{Trouble: Old Red Galaxies and Big Dusty Starbursts}

The new challenge is to go beyond global quantities like those
discussed above, and to ask what kind of galaxies host those stars and
where the star formation is occurring. For example, we know that most
of the stellar mass in the local universe is contained in old,
massive, spheroids, while most of the star formation is occuring in
smaller-mass, disk-dominated objects. The new generation of
multi-wavelength surveys is allowing us to determine whether this was
true at earlier times. The results will pose much more stringent
constraints on models.

Current indications are that the same hierarchical models that do well
at reproducing global quantities, as discussed above, do much less
well at producing the correct proportions of different types of
galaxies as a function of redshift. In particular, while the models
seem to produce \emph{enough stellar mass globally}, at redshifts of
about unity and above they do not produce as many massive/luminous
\emph{red} galaxies as are observed. For example, it has been noted in
several different studies (e.g. \cite{firth,cimatti,somerville:04})
that semi-analytic models do not produce enough `extremely red
objects' (EROs), objects with observed-frame optical-infrared colors
redder than a passively evolving elliptical (e.g. $(R-K)_{\rm VEGA} >
5$--6), which tend to be at redshifts $0.8 \la z \la 1.5$ (see
e.g. \cite{moustakas:04}). Semi-analytic models underproduce these
objects by factors of two to ten, depending on the precise color and
magnitude cut, and the specific star formation recipes used. Also,
although semi-analytic models can be tuned to give passable agreement
with the rest frame $U-V$ color-magnitude relation at $z=0$, the model
fails to produce enough luminous/massive, red, bulge-dominated
galaxies at $z\ga0.5$--1, based on measurements from COMBO-17 and the
DEEP redshift survey.

A possibly related problem is that, when absorption and emission by
dust are modeled in detail, semi-analytic models also fail to produce
enough luminous sub-mm galaxies \cite{guiderdoni,devriendt}, now
believed to be predominantly at redshifts $z\sim2.5$
\cite{chapman:03}. This problem is demonstrated in Fig.~\ref{fig:sfh}:
the dashed curve shows the star formation density needed to account
for the sub-mm sources and the Far-IR extragalactic background
\cite{blain:99,blain:02}. The required star formation at
$z\sim1.5$--2.5 greatly exceeds that predicted by the semi-analytic
models or hydro simulations. One way to solve this problem is to
assume that the IMF was very top-heavy at early times, or in
starbursts (which account for a larger fraction of star formation at
high redshift), as we heard from Cedric Lacey at this meeting. It is
possible, though, that these two problems are reflections of the same
underlying difficulty: perhaps not enough massive objects are being
created (in huge dust-enshrouded bursts) at $z\ga2$, which is why not
enough old, red, quiescent massive objects are in place at
$z\sim1$--2. However, the integrated stellar mass density at low
redshift poses a strong constraint --- it would clearly be exceeded if
the Blain et al. curve were integrated. This remains a puzzle.
 
\section{Summary}

To summarize, hierarchical models of galaxy formation (whether based
on semi-analytic or N-body+hydro techniques) are currently able to
satisfy observational constraints on the global history of star
formation and stellar mass assembly from \emph{optically selected}
galaxies, from $z\sim6$ to the present. However, they fail badly at
reproducing two very different kinds of massive objects at high
redshift: quiescent red spheroids, and vigorously star forming,
heavily dust-obscured ULIRG-like objects detected in the mid-IR and
sub-mm. It is possible, though not certain, that these problems are
connected. One can think of many physical mechanisms, neglected in the
current modeling, that could \emph{quench} star formation and produce
more red objects, but these mechanisms alone could only make galaxies
dimmer and less massive. What seems to be needed is a mechanism that
\emph{boosts} star formation but then rapidly \emph{quenches} it ---
in massive systems preferentially. This may sound like an old story
(monolithic dissipative collapse). What is needed is a \emph{physical
mechanism} that produces such a process within the framework of the
otherwise largely successful hierarchical structure formation
model. There is a growing belief that AGN-driven feedback will
probably play a major role in solving these problems, but many details
remain to be worked out. I conclude that the multi-wavelength universe
is a messy but exciting place to live.


\begin{thebibliography}{}
\addcontentsline{toc}{section}{References}
\bibitem{barger} A. J. Barger, L. L. Cowie, E. A. Richards: AJ, \textbf{119},
2092 (2000)
\bibitem{bell:03} E. F. Bell, D. H. McIntosh, N. Katz, M. D. Weinberg:
ApJS, \textbf{149}, 289 (2003)
\bibitem{blain:99} A. W. Blain, I. Smail, R. J. Ivison, J.-P. Kneib:
MNRAS, \textbf{302}, 632 (1999)
\bibitem{blain:02} A. W. Blain, I. Smail, R. J. Ivison, J.-P. Kneib,
  D. T. Frayer, PhR, \textbf{369}, 111 (2002)
\bibitem{bsf04} J.M. Bromley, R.S. Somerville, A. C. Fabian: MNRAS, in press (astro-ph/0311008)
\bibitem{chapman:03} S.C. Chapman, A.W. Blain, R.J. Ivison,
I.R. Smail: Nature, \textbf{422}, 695 (2003)
\bibitem{cimatti} A. Cimatti et al.: A\&A, \textbf{391}, L1 (2002)
\bibitem{cole:01} S. Cole et al.: MNRAS, \textbf{326}, 255 (2001)
\bibitem{daddi:00} E. Daddi, A. Cimatti, A. Renzini: A\&A,
\textbf{362}, L45 (2000)
\bibitem{daddi:03} E. Daddi et al.: ApJL, GOODS special edition, in
press (2004)
\bibitem{devriendt} J.E.G. Devriendt, B. Guiderdoni, A\&A, \textbf{363}, 851 (2000)
\bibitem{dpfb} M. Dickinson, C. Papovich, H.C. Ferguson, T. Budavari:
ApJ, \textbf{587}, 25 (2003)
\bibitem{fan:01} X. Fan et al.: AJ, \textbf{122}, 2833 (2001)
\bibitem{firth} A. E. Firth et al.: MNRAS, \textbf{332}, 617 (2002)
\bibitem{genzel:venice} R. Genzel, this meeting
\bibitem{giavalisco:04} M. Giavalisco et al.: ApJL, GOODS special
edition, in press (2004)
\bibitem{gdds} K. Glazebrook et al. (astro-ph/0401037)
\bibitem{guiderdoni} B. Guiderdoni, Hivon, Eric; Bouchet, Francois R.;
Maffei, Bruno: MNRAS, \textbf{295}, 877 (1998)
\bibitem{idzi} R. Idzi, R. S. Somerville, C. Papovich, H. C. Ferguson,
M. Giavalisco, C. Kretchmer, J. Lotz: ApJL, GOODS special edition, in
press (2004)
\bibitem{lilly} Lilly, S. J.; Le Fevre, O.; Hammer, F.; Crampton,
David: ApJL, \textbf{460}, 1 (1996)
\bibitem{madau} P. Madau, H.C. Ferguson, M. E. Dickinson,
M. Giavalisco, C. C. Steidel, A. Fruchter: MNRAS,
\textbf{283}, 1388 (1996)
\bibitem{moustakas:04} L. A. Moustakas et al.: ApJL, GOODS special
edition, in press (2004)
\bibitem{moustakas:pc} L. A. Moustakas et al., poster
\bibitem{papovich} C. Papovich, M. Dickinson, H. C. Ferguson: ApJ,
\textbf{559}, 620 (2001)
\bibitem{renzini} A. Renzini, in `The Young Universe', D'Odorico,
Fontana, Giallongo, eds, ASP Conf. Ser., Vol. Astron. Soc. Pac., San
Francisco
\bibitem{rudnick} G. Rudnick et al.: ApJ, \textbf{599}, 847 (2003)
\bibitem{shapley} A. E. Shapley, C. C. Steidel, K. L. Adelberger,
 M. Dickinson, M. Giavalisco, M. Pettini: ApJ, \textbf{562}, 95 (2001)
\bibitem{sheth-tormen} R. K. Sheth, G. Tormen: MNRAS, \textbf{308}, 119 (1999)
\bibitem{somerville:04} R.S. Somerville et al.: ApJL, GOODS special
edition, in press
\bibitem{spf} R.S. Somerville, J.R. Primack, S.M. Faber: MNRAS,
\textbf{320}, 504 (2001)
\bibitem{springel} V. Springel, L. Hernquist: MNRAS, \textbf{339}, 312
(2003)
\bibitem{steidel:99} C.C. Steidel, K. L. Adelberger, M. Giavalisco,
M. Dickinson, M. Pettini: ApJ, \textbf{519}, 1 (1999)

\end{thebibliography}
\end{document}